\documentclass{article}
\usepackage{spconf,amsmath,graphicx}
\usepackage{mathrsfs}
\usepackage{amsfonts,amssymb}
\usepackage{float}
\usepackage[colorlinks, linkcolor=black,  citecolor=black, urlcolor=blue]{hyperref}
\usepackage{etoolbox}
\patchcmd\thebibliography
{\labelsep}
{\labelsep\itemsep=-1pt\relax}
{}
{\typeout{Couldn't patch the command}}

\title{Coarse-to-fine Kidney Segmentation Framework Incorporating with Abnormal Detection and Correction}
\name{Yue Zhang$^{1,2}$, Jiong Wu$^{3}$, Yu Zhou$^{4}$, Yifan Chen$^{5}$ and Xiaoying Tang$^{1}$
	\thanks{This study was supported by the National Natural Science Foundation
		of China (NSFC 81501546) and the National Key R\&D Program of China
		(2017YFC0112404).
		Corresponding author are Prof. Yifan Chen (\url{yifan.chen@uestc.edu.cn}) and Dr. Xiaoying Tang (\url{tangxy@sustech.edu.cn})
		}}
\address{
	$^1$Department of Electrical and Electronic Engineering, Southern University of Science and Technology,\\ Shenzhen, China\\
	$^2$Department of Electrical and Electronic Engineering, The University of Hong Kong, Hong Kong, China\\
	$^3$School of Electronics and Information Technology, Sun Yat-sen University, Guangzhou, China.\\
	$^4$Beijing Institute of Collaborative Innovation, Beijing, China.\\
	$^5$School of Life Science and Technology, University of Electronic Science and Technology of China,\\ Chengdu, China\\
}
\begin{document}
%
\maketitle
\begin{abstract}
In this paper, we formulated the kidney segmentation task in a coarse-to-fine fashion, predicting a coarse label based on the entire CT image and a fine label based on the coarse segmentation and separated image patches.
A key difference between the two stages lies in how input images were preprocessed; for the coarse segmentation, each 2D CT slice was normalized to be of the same image size (but possible different pixel size), and for the fine segmentation, each 2D CT slice was first resampled to be of the same pixel size and then cropped to be of the same image size. 
In other words, the image inputs to the coarse segmentation were 2D CT slices of the same image size whereas those to the fine segmentation were 2D MR patches of the same image size as well as the same pixel size.
In addition, we design an abnormal detection method based on component analysis and use another 2D convolutional neural network to correct these abnormal regions between two stages.
A total of 168 CT images were used to train the proposed framework and the evaluations were conducted qualitatively on other 42 testing images.
The proposed method showed promising results and achieved 94.53 \% averaged DSC in testing data. 
\end{abstract}
\section{Introduction}
Kidney cancer is one of the most common types of cancer and it is estimated that more than 175,000 people dying of kidney cancer and 400,000 new cases \cite{bray2018global}. 
The morphometry of the kidney tumor revealed from contrast-enhanced Computed Tomography (CT) is an important factor in the clinical decision but it is difficult to segment tumors as they are usually small and have the similar contrast with tissues in background.
As such, techniques that can automatically and accurately segment the kidney (including tumor) from CT images is urgently needed for research and clinical purposes.
It is a necessary and important prior step to do further analysis for a kidney tumor.
\par There are a lot of challenges to segment kidney. 
For example, the kidney occupies a small part in the whole CT images and the relative size and location for two kidneys are varied, especially for kidneys with tumors.  
In the last few years, Deep learning has been widely used in natural image segmentation and have promising results \cite{badrinarayanan2017segnet}. 
However, challenges make that directly using deep learning is usually ineffective for the medical image segmentation tasks.
\par Motivated by that a smaller input region may lead to a more accurate segmentation in deep learning-based methods \cite{zhou2017fixed}, coarse to fine strategy is useful in medical image segmentation.  
Jia et al. proposed a coarse-to-fine segmentation algorithm combining the atlas-based method and convolutional neural network (CNN) \cite{jia2018atlas}. 
Its coarse segmentation stage using registration and fusion nevertheless is quite time-consuming.
For other CNN methods performing direct segmentation, they mainly used 2D or 3D patches \cite{wu2019multi}, which however still have two limitations. 
On the one hand, the image intensities of surrounding tissues are similar to that of the kidney, which may cause false positive. 
On the other hand, a large number of patches need to be extracted at the testing stage, which is again time-consuming.
Different from the patch based method, 2D sliced based can capture global shape features \cite{zhang2019Prostate}, but it usually causes abnormal missing or extra sub-regions due to the fuzzy background. 

\par In such context, we propose a novel and efficient coarse-to-fine (C2F) segmentation framework and apply it to kidney segmentation using CT images.
We formulate kidney segmentation as a two-step task involving two CNNs; one CNN is trained to predict the rough location of the kidney based on the entire 2D MR slices (coarse stage) and the other CNN is trained to predict the accurate shape based on the previously-obtained coarse segmentation and cropped 2D CT slices (fine stage). 
All of our experiments were conducted on the MICCAI KiTS19 Challenge dataset \cite{heller2019kits19}.

\section{Method}
Let a CT image be \textbf{X} and the corresponding ground truth segmentation be \textbf{Y} where $y_i =1$ indicates a foreground voxel(including kidney and tumor).
Both image size and voxel size are usually varied for the clinical dataset.
Since spatially inconsistent data might not be ideal for machine learning applications, we firstly interpolated all the dataset to the same voxel size $d\times h\times w $ but the corresponding image size $ D\times H \times W $ still varied.
Previous coarse-fo-fine work \cite{zhou2017fixed} inspires us to make use of a coarsely predicted segmentation mask to constrain the input region with the same image size and the same pixel size. 
In other words, we use one CNN (coarse segmentation model $\mathbb{C}$) to find the rough location of the kidney and then use another CNN (fine segmentation model $\mathbb{F}$) to localize the kidney more accurately.
We also design a CNN (abnormal correction model $\mathbb{A}$) to correct the ill segmentation in the coarse stage.
$\mathbb{C}$, $\mathbb{A}$ and $\mathbb{F}$ have the same network architecture and the architecture used in this work is U-net \cite{unet}.

\subsection{Training}
Firstly, we divide each 3D volume $\textbf{X}$, $\textbf{Y}$ into a set of 2D slices $\{\textbf{x}$, $\textbf{y}\}$ with image size $H\times W$ and pixel size $h \times w$. 
The 2D slices are typically isotropic in terms of both size and resolution, i.e., $H=W$ and $h=w$.
Otherwise, resizing and padding can be easily used to make the 2D slices isotropic.
For 2D slices of different subjects, either the image size or the pixel size can be different.

To begin our C2F segmentation pipeline, we conduct two-fold preprocessing.
On the one hand, we directly resize all $\{\textbf{x}$, $\textbf{y}\}$ to be of the same image size $H^\mathbb{C}\times W^\mathbb{C}$ (this transformation is denoted as $\mathcal{R}_{W}$) and obtain 2D image and ground truth pairs $\{\textbf{x}^\mathbb{C},\textbf{y}^\mathbb{C} \}$, which are then fed into a coarse segmentation model $\mathbb{C}:\textbf{p}^\mathbb{C} = f(\textbf{W}^\mathbb{C},\textbf{x}^\mathbb{C})$, where $\textbf{W}^\mathbb{C}$ denotes the model parameters and $\textbf{p}^\mathbb{C}$ is the predicted mask at the coarse stage. 
On the other hand, we crop images with the same pixel size ($w\times h$ mm$^2$) to obtain image patches of the same image size $W^\mathbb{F} \times H^\mathbb{F}$ surrounding a single kidney (this transformation is denoted as $\mathcal{C}^\mathbb{F}$). 
The processed images and labels $\{\textbf{x}^\mathbb{F},\textbf{y}^\mathbb{F}\}$ are then fed into a fine segmentation model $\mathbb{F}:\textbf{p}^\mathbb{F} = f(\textbf{W}^\mathbb{F},\textbf{x}^\mathbb{F})$. 
To train an abnormal correction model $\mathbb{A}$, the similar cropping operation $\mathcal{C}^\mathbb{A}$ was used along the sagittal plane to get image patches with the same image size ($D^\mathbb{A} \times H^\mathbb{A}$) as well the same pixel size ($d\times h$ mm$^2$). 

In the context of a deep segmentation network, the Dice loss $ \mathcal{L}(\textbf{p},\textbf{y})$ is optimized with respects to $\textbf{W}$ via gradient back-propagation. The objective function is
\begin{equation}
\textbf{W}^*=\mathop{\arg}\min_{\textbf{W}} \frac{1}{N} \sum_{n=1}^{N} \mathcal{L}(\textbf{p}_n,\textbf{y}_n),  
\end{equation} 
where $N$ denotes the total number of samples, $\textbf{W}^*$ denotes the optimal weights obtained from the training procedure.
After the training process, $\textbf{W}^\mathbb{*C}$, $\textbf{W}^\mathbb{*A}$ and $\textbf{W}^\mathbb{*F}$ are saved.

\subsection{Testing}
\begin{figure*}[htbp]
	\centering
	\includegraphics[scale=0.16]{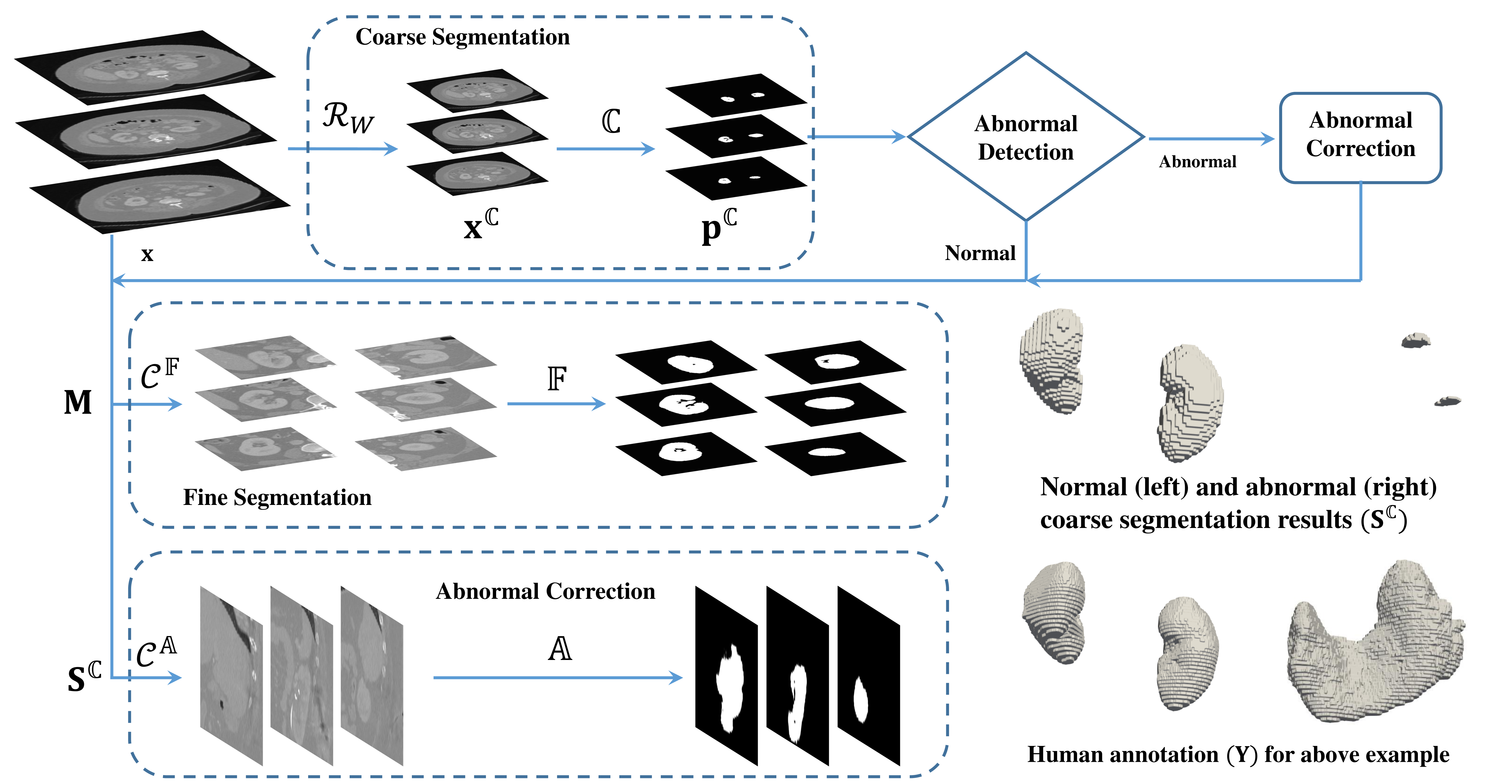}
	\caption{Flow chart of the entire testing procedure }
	\label{fig:test}
\end{figure*}
The overall testing flow chart is shown in Fig. \ref{fig:test} and the voxel spacing of input images has been normalized to be $d \times h\times w$ as that in the training stage. 
The image transformation operations $\mathcal{R}_{W}$ and cropping operation $\mathcal{C}^\mathbb{F}$ and $\mathcal{C}^\mathbb{A}$ are similar to that used in the training stage.
The only difference is that the cropping operation refers to the ground truth in the training stage now refers to the coarse segmentation in the testing stage.  
$\mathcal{R}_{W}^{-1},\mathcal{C}^\mathbb{-F}$ and $\mathcal{C}^\mathbb{-A}$ respectively denote their inverse transformation operations.

\subsubsection{Coarse-to-fine prediction}
At the coarse segmentation stage, we resize all 2D CT slices $\textbf{x}$ at the axial view to be of image size $H^\mathbb{C}\times W^\mathbb{C}$ and predict the coarse segmentation result $\textbf{S}^{\mathbb{C}}$ by
\begin{equation}
\begin{aligned}
\textbf{p}^{\mathbb{C}} &= f(\textbf{W}^\mathbb{C*},\mathcal{R}_{W} (\textbf{x})) \\
\textbf{S}^{\mathbb{C}} &= \{ \mathcal{R}_{W}^{-1}(\textbf{p}^{\mathbb{C}} )\},
\end{aligned}
\end{equation} 
where $\{ \cdot \}$ denote composing all axial sliced segmentation results $\textbf{p}^\mathbb{C}$ of one 3D coarse segmentation $\textbf{S}^{\mathbb{C}}$ for each 3D CT image $\textbf{X}$.

And then $\textbf{S}^{\mathbb{C}}$ will be judged according to the criterion defined in \ref{abnormal}.
Under different circumstances, the mask $\textbf{M}$ used to guide fine segmentation stage is defined by
\begin{equation}
\textbf{M}=\left\{
\begin{aligned}
&\textbf{S}^{\mathbb{C}}, & \rm Normal\\
&\{   
\mathcal{C}^{-\mathbb{A}}(
f(\textbf{W}^\mathbb{A*},  \mathcal{C}^\mathbb{A} (\textbf{x},  \textbf{S}^{\mathbb{C}} ))
)
\}	, & \rm Abnormal.\\
\end{aligned}
\right. 
\end{equation} 
For normal case, $\textbf{S}^{\mathbb{C}}$ is directly set as $\textbf{M}$.
For abnormal case, $\textbf{S}^{\mathbb{C}}$ is used to decide the centroid in the sagittal plane and the cropping operation $\mathcal{C}^\mathbb{A}$ is used to get image patches with size $D^\mathbb{A} \times H^\mathbb{A}$. 
$\mathcal{C}^\mathbb{-A}$ is used to pad the predication to the original size using 0.

At the fine segmentation stage, we crop the resultant images to obtain image patches of image size $H^\mathbb{F}\times W^\mathbb{F}$ according to the separated centroid in axial plane decided by $\textbf{M}$.
Two kidneys are predicted separately in fine segmentation stage by 
\begin{equation}
\textbf{S}^{\mathbb{F}}=
\{   
\mathcal{C}^{-\mathbb{F}}(
f(\textbf{W}^\mathbb{F*},  \mathcal{C}^\mathbb{F} (\textbf{x},  \textbf{M} ))
)
\}.
\end{equation} 
\subsubsection{Abnormal detection}\label{abnormal}
A typical drawback of coarse-to-fine strategy is that the performance of the fine model depends on that of the coarse model. 
For example, most people have two kidneys, but there are a few who only have one kidney.
As shown in Fig. \ref{fig:test}, the abnormal coarse segmentation result fails to detect the whole kidney.

For these abnormal cases, we design an automatic abnormal detection method based on component analysis and correct it using a CNN.
Specifically, we first extract all sub structures of the coarse segmentation result $\textbf{S}^{\mathbb{C}}$ using connected component analysis \cite{van2014scikit}.
The voxel number of each substructure can be counted.
Given that kidney volumes are 202 $\pm$ 36 ml for men and 154 $\pm$ 33 ml for women \cite{cheong2007normal} and we have normalized all image to the same voxel size at the beginning, we can set a threshold voxel number $TH_{vn}$ to count the kidney number $N_{kidney}$.
Based on $N_{kidney}$ detected in $\textbf{S}^{\mathbb{C}}$, we define a discriminate criterion like 
\begin{equation}
\left\{
\begin{aligned}
&N_{kidney}=2, & \rm Normal\\
&\rm else	, & \rm Abnormal.\\
\end{aligned}
\right. 
\end{equation} 
\subsection{Implementation Details}
The normalized spacing size $d\times h\times w$ is set as $3\times0.7816\times0.7816$ mm$^3$ as the public dataset has the interpolated version with this voxel spacing.
The normalized image size in three CNN models ($\mathbb{C},\mathbb{A},\mathbb{F}$) are set as $H^\mathbb{C}=W^\mathbb{C}=128$, $H^\mathbb{A}=256$, $D^\mathbb{A}=64$, and $H^\mathbb{F}=W^\mathbb{F}=160$ empirically.
The threshold voxel number $TH_{vn}$ is set as 10000, which is roughly equal to 18 ml based on previous setting voxel spacing.
We set this value relative lower to the real kidney volume to reduce the false-negative rate and this value is large enough to remove the noisy small lesions and judge the predicted kidney number.

\section{EXPERIMENTS AND RESULTS}
\subsection{Dataset}
All data used in this study came from the MICCAI KiTS19 Challenge \cite{heller2019kits19}. 
The public dataset consists of 210 abdominal CT images and the associated segmentation ground truth. 
And we divided them into 168 training data and 42 testing data
The testing data were identified to be the images of indices $\{0,5,10,15,\dots,205 \}$.

\begin{figure}[htbp]
	\centering
	\includegraphics[scale=0.33]{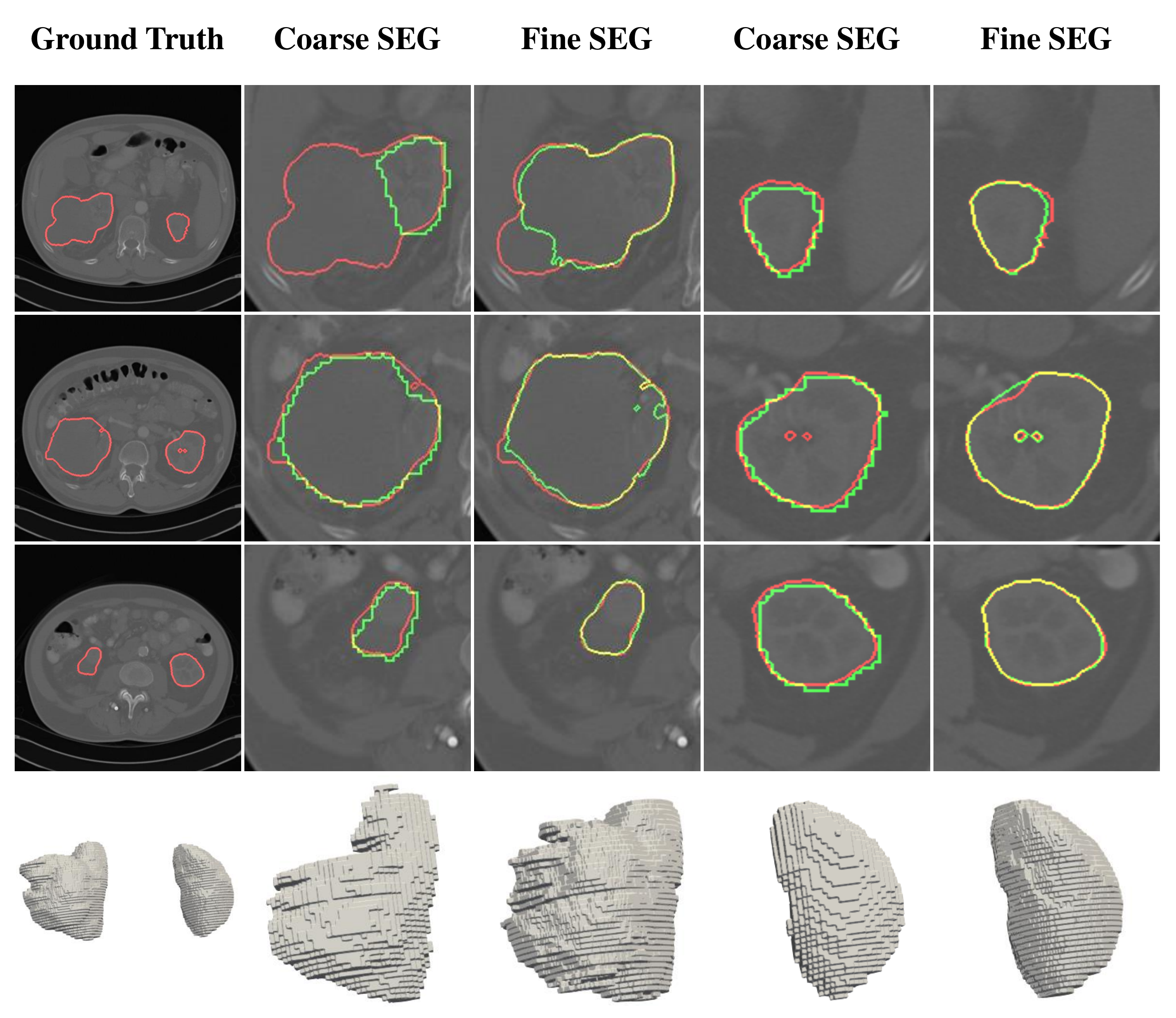}
	\caption{Representative segmentation results with human annotation.
		The red and green curves are manual delineation, segmentation in coarse stage and fine stage, respectively.
		Segmentation boundaries of our fine stage segmentation results and ground truth sometimes are too close to distinguish.}
	\label{fig:results}
\end{figure}

\subsection{Results}
The segmentation results of the proposed approach are shown in Fig. \ref{fig:results}.
The coarse segmentation shows zigzag edge but the fine segmentation is smooth and sometimes totally the same as ground truth.
These zigzag edge can be partially repaired if we increase the image size $H^\mathbb{C}=W^\mathbb{C}$ in the coarse model.
But the missing parts shown in the first row will still damage the performances. 
The 3D reconstruction of ground truth and segmentation are also shown in Fig. \ref{fig:results}.
Compared with human annotation, the fine segmentation can keep the overall shape and more details than coarse segmentation.

\par Volumetric Dice similarity coefficient (vDSC) is a major quantitative standard to compare the similarity between ground truth and segmentation results, 
i.e.,  the higher the DSC is, the better the segmentation results will be.
Comparisons between the coarse segmentation and fine segmentation in terms of the average with standard deviation, max, and min DSC scores over 42 testing images are tabulated in Table \ref{table1}.
Evidently, the fine stage segmentation has the highest segmentation accuracy and the lowest standard deviation. 

\begin{table}[H]
	\centering
	\vspace{-2em}
	\footnotesize    
	\renewcommand{\arraystretch}{1.2}   
	\setlength{\abovecaptionskip}{0ex}%
	\setlength{\belowcaptionskip}{2pt}%
	\caption{Quantitative comparisons of coarse segmentation and fine segmentation. Keys: STD--Standard deviation.}
	\medskip
	\label{table1}
	\begin{tabular}{cccc}  \hline
		&  Mean $\pm$ STD [\%] &Max [\%] & Min [\%] \\ \hline
		Coarse &  84.47 $\pm$ 14.70 &92.75  & 2.36\\
		Fine &  \textbf{94.53} $\pm$ \textbf{8.33} &\textbf{98.69}  & \textbf{57.89}\\
		\hline
	\end{tabular}
\end{table}

\section{CONCLUSION}
For a sliced based CNN, the input image size should be the same and the image size of the 2D slice is usually varied. 
In this work, we proposed a coarse-fo-fine framework jointly considering image size and pixel size.
In addition, we designed an abnormal detection and correction method between coarse stage and fine stage, 
which was efficient to repair the abnormal coarse segmentation and guarantee better performance in the fine stage.
\bibliographystyle{IEEEbib}

\end{document}